\documentclass[10pt,letterpaper]{article}
\usepackage[top=0.85in,left=2.75in,footskip=0.75in]{geometry}

\usepackage{amsmath,amssymb}

\usepackage{changepage}

\usepackage[utf8x]{inputenc}

\usepackage{textcomp,marvosym}

\usepackage{cite}

\usepackage{nameref,hyperref}

\usepackage{microtype}
\DisableLigatures[f]{encoding = *, family = * }

\usepackage[table]{xcolor}

\usepackage{array}

\newcolumntype{+}{!{\vrule width 2pt}}

\newlength\savedwidth

\usepackage{tabularx}
\newcolumntype{C}[1]{>{\hsize=#1\hsize\centering\arraybackslash}X}%
\newcolumntype{L}[1]{>{\hsize=#1\hsize\raggedright\arraybackslash}X}%
\usepackage{multirow}
\usepackage{graphicx}
\usepackage[caption=false,font=normalsize,labelfont=sf,textfont=sf]{subfig}

\usepackage{xspace}

\newcommand*{\ie}{\textit{i.e}.\@\xspace}

\makeatletter
\newcommand*{\etc}{%
    \@ifnextchar{.}%
        {\textit{etc}}%
        {\textit{etc}.\@\xspace}%
}
\makeatother

\raggedright
\usepackage[aboveskip=1pt,labelfont=bf,labelsep=period,justification=raggedright,singlelinecheck=off]{caption}

\makeatletter
\renewcommand{\@biblabel}[1]{\quad#1.}
\makeatother

\usepackage{lastpage,fancyhdr,graphicx}
\usepackage{epstopdf}
\fancyheadoffset[L]{2.25in}
\fancyfootoffset[L]{2.25in}
\begin{document}
\vspace*{0.2in}

\begin{flushleft}
{\Large
\textbf\newline{Automated Grading System of Retinal Arterio-venous Crossing Patterns: A Deep Learning Approach Replicating Ophthalmologist's Diagnostic Process of Arteriolosclerosis} 
}
\newline
\\
Liangzhi Li\textsuperscript{1},
Manisha Verma\textsuperscript{1},
Bowen Wang\textsuperscript{1},
Yuta Nakashima\textsuperscript{1},
Hajime Nagahara\textsuperscript{1},
Ryo Kawasaki\textsuperscript{2}*
\\
\bigskip
\textbf{1} Institute for Datability Science (IDS), Osaka University, Osaka 565-0871, Japan
\\
\textbf{2} Graduate School of Medicine, Osaka University, Osaka 565-0871, Japan
\\
\bigskip

%
%









* ryo.kawasaki@ophthal.med.osaka-u.ac.jp

\end{flushleft}
\section*{Abstract}

\textbf{Background and Objective}: The morphological \textcolor{black}{feature} of retinal arterio-venous crossing patterns is a valuable source of cardiovascular risk stratification as it directly captures vascular health. \textcolor{black}{Although Scheie's classification, which was proposed in 1953, has been used to grade the severity of arteriolosclerosis as diagnostic criteria}, it is not widely used in clinical \textcolor{black}{settings} as mastering this grading is challenging as it requires vast \textcolor{black}{experience}. In this paper, we propose a deep learning approach to replicate a diagnostic process of ophthalmologists while providing a \textcolor{black}{checkpoint} to secure explainability to understand the grading process.

\textbf{Methods}:  
The proposed pipeline is three-fold to replicate a diagnostic process of ophthalmologists. First,  
we adopt segmentation and classification models to automatically obtain vessels in a retinal image with the corresponding artery/vein labels and find candidate arterio-venous crossing points. Second, we use \textcolor{black}{a} classification model to validate the true crossing point. At last, the grade of severity for the vessel crossings is classified. 
To better address the problem of label ambiguity and imbalanced label distribution, we propose a new model, named multi-diagnosis team network (MDTNet), in which the sub-models with different structures or different loss functions provide different decisions. MDTNet unifies these diverse theories to give the final decision with high accuracy.

\textbf{Results}:
Our automated grading pipeline was able to validate crossing points with precision and recall of $96.3\%$ and $96.3\%$, respectively. Among correctly detected crossing points, the kappa value for the agreement between the grading by a retina specialist and the estimated score was $0.85$, with an accuracy of $0.92$.
The numerical results demonstrate that our method can achieve a good performance in both \textcolor{black}{arterio-venous} crossing validation and severity grading tasks following the diagnostic process of ophthalmologists.

\textbf{Conclusions}: By the proposed models, we could build a pipeline reproducing \textcolor{black}{ophthalmologists' diagnostic process without} requiring subjective feature extractions. (The code is available for reproducibility\footnote{The code is available at \url{https://github.com/conscienceli/MDTNet}}.)

\section*{\textcolor{black}{Author Summary}}
Assessment of arterio-venous crossing points in retinal images provides rich cues for quick screening of arteriosclerosis and even for classifying them into different severity grades. Considering the ever-increasing demand for ophthalmologic examination, computer-aided diagnosis (CAD) is extremely helpful for quick screening. However, retinal image analysis for CAD is a challenging task due to the high complexity of the vessel system and huge visual differences among retinal images.
To address the aforementioned problems, we propose a whole pipeline for an automatic method for severity grading of artery hardening. Our method can find and validate possible arterio-venous crossing points, for which the severity grade is predicted. We also design a new model, MDTNet, which uses the focal loss to address the problem of data ambiguity and unbalance. Therefore, we believe that this research contributes to the advancement of research in machine learning on retinal images.



\section*{Introduction}
Retina provides a window to directly visualize vascular structure \textit{in vivo}, and ophthalmologic examination has been regarded as an important routine for detecting not only eye diseases but also ocular manifestations of cardiovascular diseases or \textcolor{black}{their} accumulated risks \cite{chatziralli2012value}. Among these detectable retinal vascular signs, arteriolosclerosis is critical yet asymptomatic, of which diagnosis requires detailed retinal observation. It is not widely conducted in the \textcolor{black}{modern} medical practice as it depends on mostly subjective qualitative observations, and most importantly, it requires vast experiences. 

\begin{figure}[t]
  \centering
  \includegraphics[width=1\textwidth]{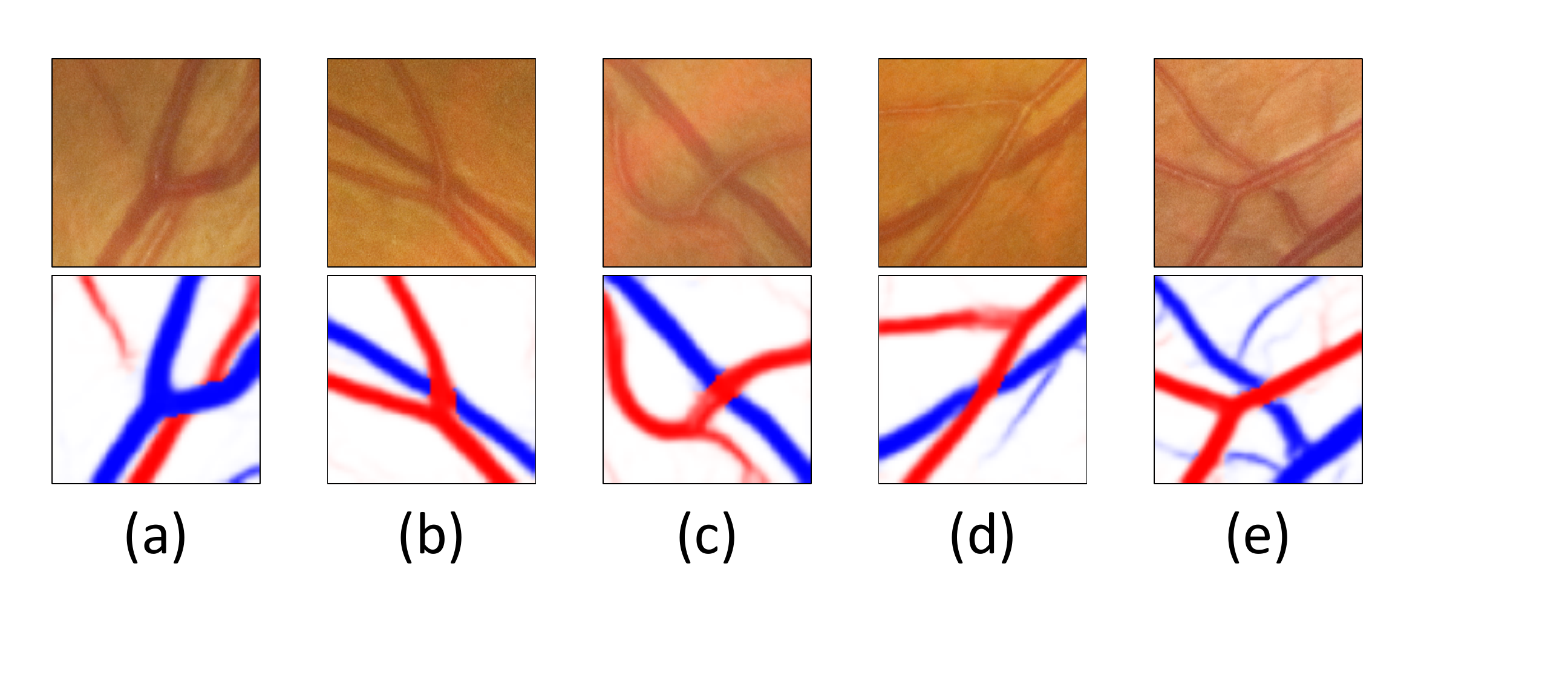}
  \caption{Typical examples of our prediction targets. Images in the first and second rows are raw retinal patches and automatically-generated vessel maps with manually-annotated artery/vein labels, respectively. Red represents arteries while blue represents veins. (a) is false crossing (the vein runs above the artery), while (b)--(e) are for \textit{none}, \textit{mild}, \textit{moderate}, and \textit{severe} grades, respectively. Note that even the state-of-the-art segmentation techniques cannot capture caliber narrowing, therefore, the arterioloscleroses are not very obvious in the vessel maps.}
  \label{fig_story}
\end{figure}

Assessment of arterio-venous crossing points in retinal images provides rich cues \textcolor{black}{for screening} arteriosclerosis and for evaluating accumulated cardiovascular risks. Typically, arterio-venous crossing points are classified into severity grades \cite{hubbard1999methods}. The assessment is based on some diagnostic criteria, for example, Scheie's classification \cite{WALSH19821127}, as shown in Figs.~\ref{fig_story}(b)--(e). The grades are described as follows: (i) \textit{none} (no anomaly observed); (ii) \textit{mild} (slight shrink in the caliber at venular edges);  (iii) \textit{moderate} (narrowed caliber at a single venular edge); and (iv) \textit{severe} (narrowed caliber at both venular edges).

However, human graders are subjective and usually with different levels of \textcolor{black}{experience}, and there has been a criticism \textcolor{black}{of} the low reproducibility of severity grading, which makes grading results from human graders unreliable for clinical practice, screening, and clinical trials \cite{6547196}. Also, considering the ever-increasing demand for ophthalmologic examination, computer-aided diagnosis (CAD) is extremely helpful for quick screening. Yet, retinal image analysis for CAD is a challenging task due to the high complexity of the vessel system and huge visual differences among retinal images.

In fact, most researchers in this area have been focusing on preliminary tasks, such as vessel segmentation \cite{7042289,8036917,8341481}, artery/vein classification \cite{HUANG2018197,10.1007/978-3-319-93000-8_71,8055572}, etc. A few works address higher-level tasks \cite{hatanaka2011automatic,6547196}, mostly on top of vessel segmentation, such as vessel width measurement, vessel-to-vessel ratio calculation, etc. However, they usually struggle in actual diagnoses: Firstly, vessel segmentation in retinal images \textit{per se} is a challenging task. The vessel maps in Fig.~\ref{fig_story}(c)--(e), which are produced by the state-of-the-art segmentation model \cite{li2019iternet}, cannot capture such deformation. This may imply that deformation is too minor to be captured by segmentation models, although such kind of segmentation-based \textcolor{black}{approach} is a typical solution for automatic severity grading. Secondly, the existing methods detect arterio-venous crossing points by applying some morphological operators to vessel maps \cite{cambocombined}. This approach may not be accurate enough to find crossing points that satisfy diagnostic requirements. For example, we can only use crossing points at which the artery is above the vein for diagnosis, and Fig.~\ref{fig_story}(a) is not a diagnostic crossing point since the artery goes below the vein. 

Instead of fully relying on segmentation results, we propose a multi-stage approach, in which segmentation results are used only for finding crossing point candidates, and actual prediction of the severity grade is \textcolor{black}{conducted} for an image patch around each crossing point after validating if the crossing point is an actual and informative one. To the best of our knowledge, this is the first work proposing a fully-automatic methodology aiming at grading arteriolosclerosis through the joint detection and analysis of retinal crossings. 

Another issue in our severity grading task, which is very common in medical imaging, is the imbalanced label distribution. Most patients in our dataset have the slightest signs (\textit{none} and \textit{mild}) of arteriolosclerosis while only a few patients suffer from the \textit{severe} grades of artery hardening. Also, the boundaries among different severity labels are not always obvious, making accurate diagnosis challenging. 

Inspired by the concept of the multidisciplinary team \cite{taylor2010multidisciplinary}, which strives to make a comprehensive assessment of a patient, we propose a multi-diagnosis team network (MDTNet) in this paper to address the imbalanced label distribution and label ambiguity problems at the same time. MDTNet can combine the features from multiple classification models with different structures or different loss functions. Some of the underlying models in MDTNet use the class-balanced focal loss \cite{FocalLoss} to handle hard or rare samples, of which the original version requires hyperparameter tuning, while MDTNet can utilize the advantage of the focal loss without tuning its hyperparameters.

Our main contribution is two-fold: 
	(i) We propose a whole pipeline for an automatic method for severity grading of artery hardening. Our method can find and validate possible arterio-venous crossing points, for which the severity grade is predicted. 
	\textcolor{black}{(ii) We design a new model, MDTNet, which uses the focal loss to address the problem of data ambiguity and unbalance. 
	}

\section*{Dataset}
\textbf{Ethics Statement: }
This study was performed in accordance with the World Medical Association Declaration of Helsinki. Patients gave written informed consent to participate and the study protocol was approved by the institutional review board of the Osaka University Hospital.


\textcolor{black}{We built a vessel crossing point dataset extracted from our retinal image database of the Ohasama study, a cohort to study cardiovascular diseases risk, where we could utilize $1,440$ images in the size of $5,184 \times 3,456$ pixels, which are captured by the CR-2 AF Digital Non-Mydriatic Retinal Camera (Canon, Tokyo) between 2013 and 2017 as JPEG files. 
This database includes the medical data of $684$ people, which are with an average age of $64.5$ (standard deviation: $6.1$). The ratio between female and male is $65.2\%:34.8\%$ and $47.6\%$ of all participants have hypertension. Details of the study profile were published elsewhere \cite{inoue2009stroke}.}

\begin{figure}[t]
  \centering
  \includegraphics[width=1\textwidth]{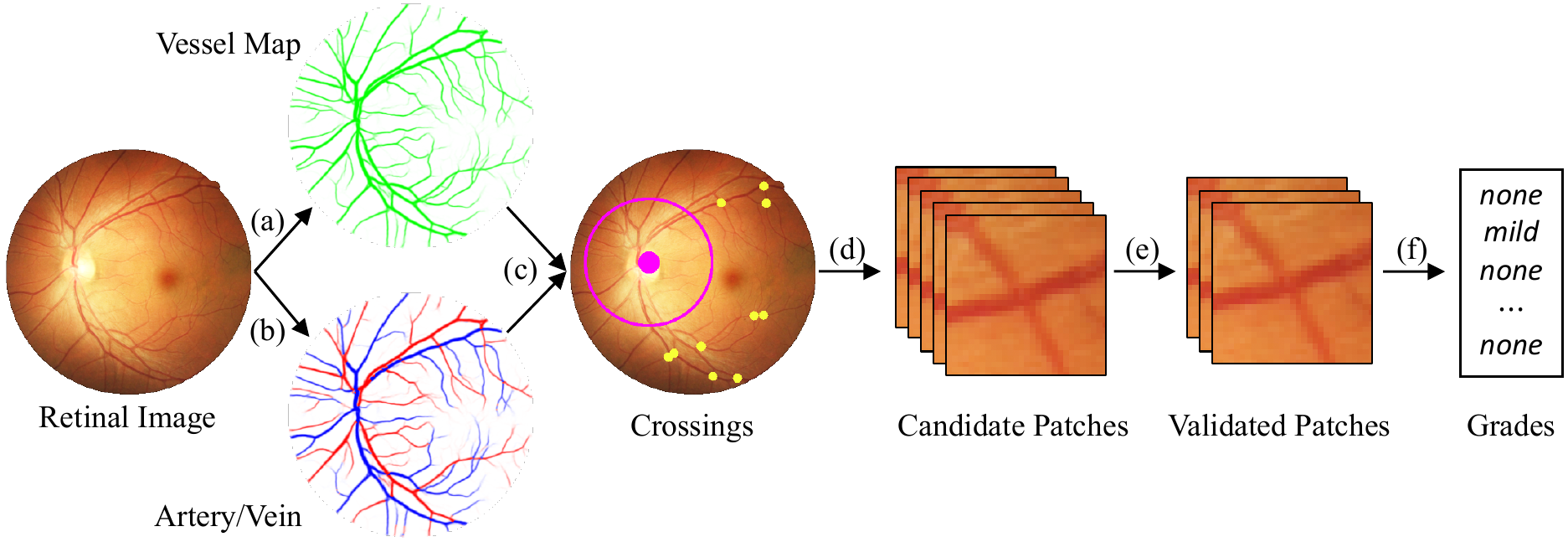}
  \caption{Overall pipeline of our severity grading.}
  \label{fig_method}
\end{figure}

To find crossing points in these images (Fig.~\ref{fig_method}(a)--(d)), we used a segmentation model (\cite{li2019iternet}) to get vessel maps. We then classified each pixel on extracted vessels into artery/vein using \cite{li2020joint}. We combine the vessel segmentation and classification results to find crossing points because classification results, which are more beneficial for crossing point detection, tend to have more errors while segmented vessel maps are more accurate. Therefore, we refine the classification results based on the vessel maps. A classic approach then finds crossing points in these refined artery/vein maps. Specifically, we find the artery pixels neighbouring vein pixels and check whether it is a crossing point or not using the skeletonized vessel map. The points marked in yellow in Fig.~\ref{fig_method} are detected crossing point candidates. Note that for cup zones as indicated by a pink circle and dot in Fig.~\ref{fig_method}, we exclude candidates because the vessel system in this area is with high complexity and thus segmentation and classification are not reliable. Image patches are of size $150\times150$, centered at the crossing point candidates.
Consequently, we detected $4,240$ crossing points and extracted corresponding image patches, centered at these crossing points. 

Each image patch was carefully reviewed by a highly experienced ophthalmologist.
Due to the errors in vessel segmentation and artery/vein classification, the detected crossing points may not be actual \textcolor{black}{or} informative. Therefore, the specialist first annotated each image patch with a label on its validity, \ie, if the image patch contains an actual and informative crossing point (\textit{true}) or not (\textit{false}). The numbers of true and false crossing points are $2,507$ and $1,733$, respectively. For each true crossing point, the specialist gave its severity label in $C = \{\mbox{\textit{none}, \textit{mild}, \textit{moderate}, \textit{severe}}\}$. The numbers of image patches with respective labels are $1,177$, $816$, $457$, and $57$. 
\textcolor{black}{In both tasks}, the datasets will be divided into training, validation, and test \textcolor{black}{sets} following a ratio of $8$:$1$:$1$. As an \textcolor{black}{examinee} may have multiple retinal images, it is important to strictly put them into one same subset to prevent the training data contamination.

\section*{Severity Grading Pipeline}

Our method forms a pipeline with three main modules, \ie, preprocessing, patch validation, and severity grade prediction. The whole pipeline is shown in Fig.~\ref{fig_method}. 

\noindent\textbf{Preprocessing} Steps (a)--(d) in the figure are preprocessing, in which the same processes as our dataset construction are applied to get image patches of $150\times150$ pixels with crossing point candidates. 

\noindent\textbf{Crossing Point Validation} 
Both crossing point validation and severity grading are classification problems, whereas validation is easier because the label distribution is more balanced and the differences between real and false crossing points are more obvious. We find that commonly used classification models, such as \cite{resnet,densenet,inception}, work well for our validation task (refer to Section \ref{sec:exp}). 

\noindent\textbf{Severity Grade Prediction} The severity grade prediction task is much more challenging: Firstly, the label distribution is highly biased. For example, samples with the \textit{none} label account for $68\%$ of the total samples, while ones with the \textit{severe} \textcolor{black}{label} only take up $3\%$. Secondly, the difference among samples with different labels may not be clear enough. Even medical doctors may make diverse decisions on a single image patch.

For such classification tasks with ambiguous or imbalanced classes, the focal loss \cite{FocalLoss} has been used, which makes a model more aware of hard samples than easy ones. 
The focal loss introduces a hyperparameter $\gamma$, on which a model's performance depends significantly. Tuning this hyperparameter is extremely important yet computationally expensive \cite{AutomatedFocalLoss}. A greater $\gamma$ may make the model focus too much on hard samples, spoiling the accuracy \textcolor{black}{of} other samples, while a smaller $\gamma$ may decrease its ability to classify hard samples. 

\begin{figure}[t]
  \centering
  \includegraphics[width=1\textwidth]{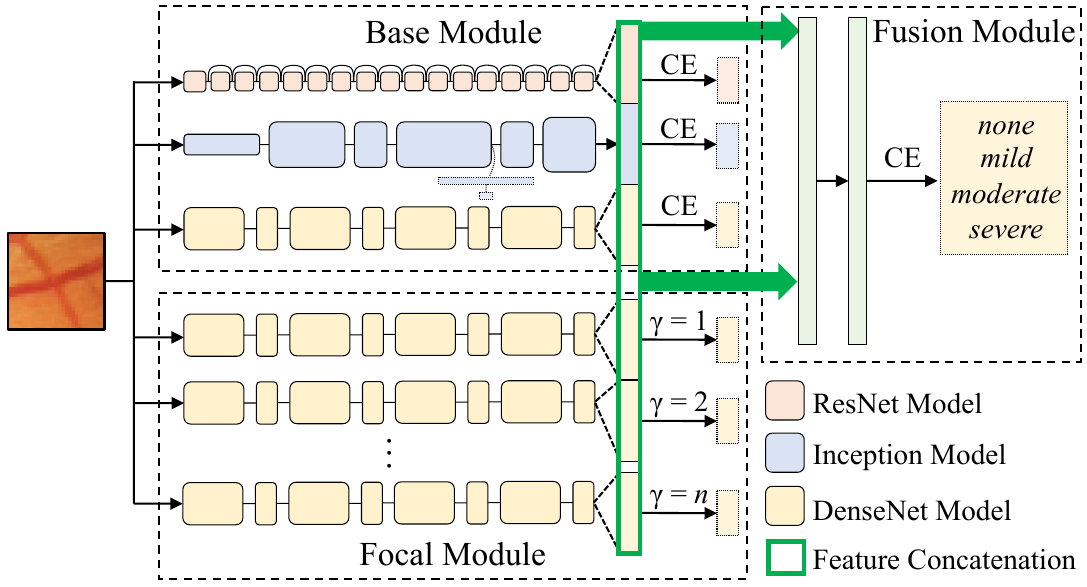}
  \caption{MDTNet for severity grade prediction.}
  \label{fig_model}
\end{figure}

We propose a multi-diagnosis team network (MDTNet) to address the aforementioned problems in severity grade prediction. As shown in Fig.~\ref{fig_model}, MDTNet consists of three modules, \ie, a base module, a focal module, and a fusion module.

\begin{figure}[t]
  \subfloat[]{\includegraphics[width=0.235\textwidth]{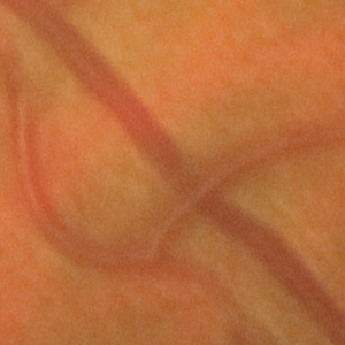}}
  \hfill
  \subfloat[]{\includegraphics[width=0.235\textwidth]{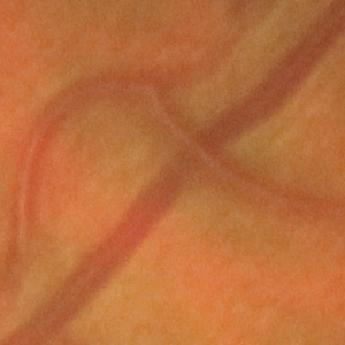}}
  \hfill
  \subfloat[]{\includegraphics[width=0.235\textwidth]{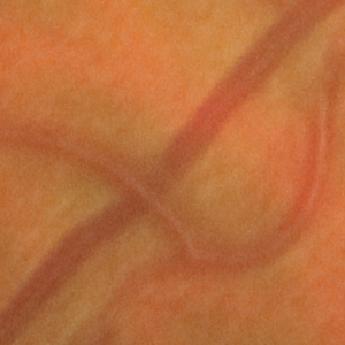}}
  \hfill
  \subfloat[]{\includegraphics[width=0.235\textwidth]{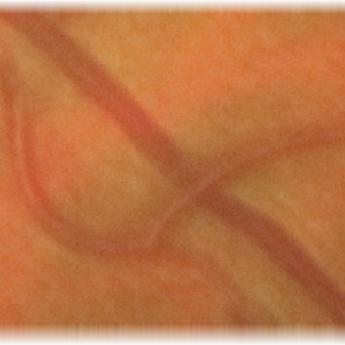}}
  \hfill
  \subfloat[]{\includegraphics[width=0.235\textwidth]{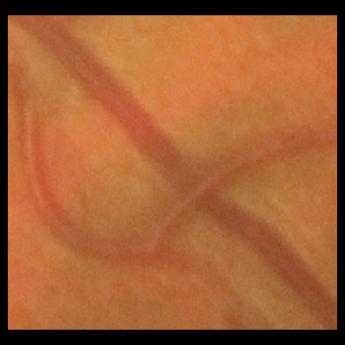}}
  \hfill
  \subfloat[]{\includegraphics[width=0.235\textwidth]{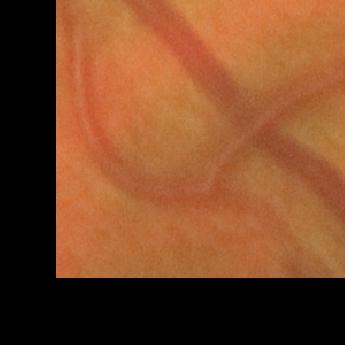}}
  \hfill
  \subfloat[]{\includegraphics[width=0.235\textwidth]{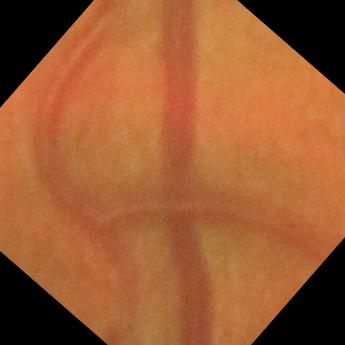}}
  \hfill
  \subfloat[]{\includegraphics[width=0.235\textwidth]{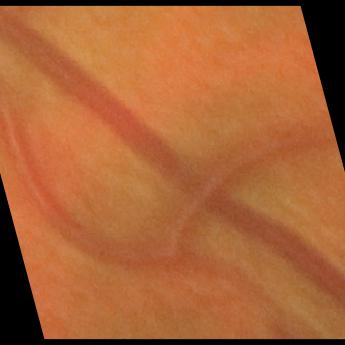}}
  \hfill
  \subfloat[]{\includegraphics[width=0.235\textwidth]{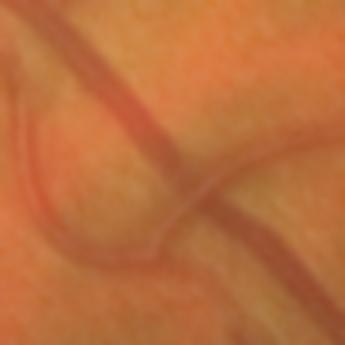}}
  \hfill
  \subfloat[]{\includegraphics[width=0.235\textwidth]{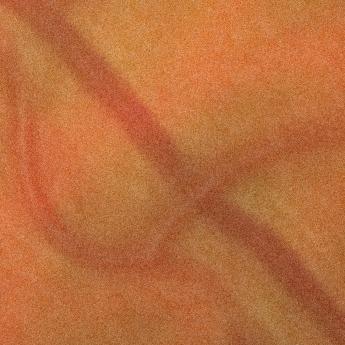}}
  \hfill
  \subfloat[]{\includegraphics[width=0.235\textwidth]{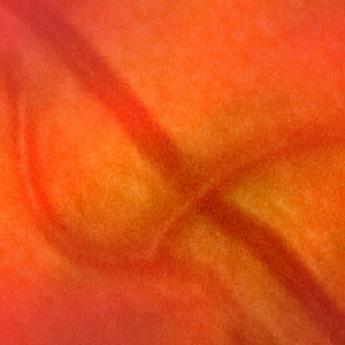}}
  \hfill
  \subfloat[]{\includegraphics[width=0.235\textwidth]{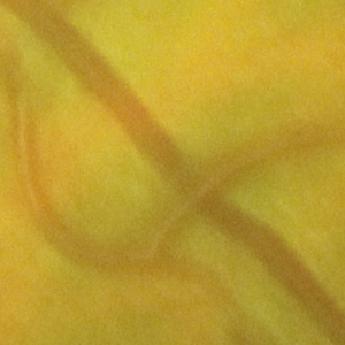}}
  \caption{Our data augmentation operator pool. (a) Raw image, (b) vertical flipping, (c) horizontal flipping, (d) cropping and padding, (e) scaling, (f) translating, (g) rotating, (h) sheering, (i) blurring, (j) additional noise, (k) additional frequency noise, and (l) color modification.}
  \label{fig_data_augmentation}
\end{figure}

The base and focal modules have multiple sub-models, and all of them take the same image patch as input. The difference between the sub-models in the base and focal modules is the losses: Ones in the base module adopt the cross entropy (CE) loss while ones in the focal module use the focal loss. These sub-models are trained independently with respective losses. The fusion module concatenates all features (\ie, the outputs of the second last layers of the sub-models) into a single vector, which is then fed into two fully-connected layers to make the final prediction. 

The focal loss is originally designed for object detection \cite{FocalLoss}, defined as
\begin{equation}
L(y, t) = -\sum_{l} t_{l}(1-y_l)^{\gamma}\log y_l,
\end{equation}
where $t$ is the one-hot representation of label and $y$ is the softmax output from a model ($t_l$ and $y_l$ are the $l$-th entries of $t$ and $y$); $\gamma$ is a hyperparameter to weight hard examples. The focal loss reduces to the CE loss when $\gamma=0$, and a larger $\gamma$ weights more on hard examples. One possible criticism of the focal loss is its sensitivity to $\gamma$. We therefore propose to ensemble sub-models with different $\gamma$'s. The hypothesis behind this choice is that different $\gamma$'s may rely on different cues for prediction and aggregating respective features may help in improving the final decision. This is embodied in the focal module. The same idea can also be applied \textcolor{black}{to} different network architectures, embodied in the base module. These sub-models thus provide diagnostic features that may complement each other. 

To cope with the imbalanced class distribution, we adopt class weighting \cite{7780949,Cui_2019_CVPR}. We multiply weight $\alpha_l = \ln N_l/\ln N$ to each term (\ie different $l$'s) in the CE/focal loss, where $N$ and $N_l$ are the numbers of all samples and of samples with the label corresponding to the $l$-th entry of $t$. We pre-train the sub-models using their own classifiers and losses, and then freeze their weights to train the additional two fully-connected layers for the final decision.

\noindent\textbf{Data Augmentation}
We adopt extensive data augmentation. During the training process, the input images have 50\% chance \textcolor{black}{of getting} each operator in Fig.~\ref{fig_data_augmentation}. Among them, (b$\sim$h) are used for shape modification, changing the locations and the shapes of the attention areas of the deep learning models; (i$\sim$k) are to provide variety on imaging quality by blurring or adding random noises; (l) represents sensor characteristics of color (hue and saturation).


\section*{Experiments and Results}\label{sec:exp}


\noindent\textbf{Implementation}
For sub-models in the base module, we used ResNet \cite{resnet}, Inception \cite{inception}, and DenseNet \cite{densenet}. In the focal module, DenseNet with $\gamma =$ $1$, $2$, or $3$ were used. All these models are \textcolor{black}{pre-trained} over the ImageNet dataset \cite{ILSVRC15}. The fully-connected layers in the fusion module are followed by the ReLU nonlinearity. For optimization, Adam \cite{adam} was adopted with a learning rate of $0.0001$.
\textcolor{black}{Models are trained on the training set, and the weights with the highest performance on the validation set are selected as the best models, which will be evaluated on the test set.}

\noindent\textbf{Performance of Base Models}
We first evaluated the performance of the base module's sub-models for the crossing point validation and severity grade prediction tasks. For comparison, we also give the results of models without pre-training (w/o PT) and without data augmentation (w/o DA), as well as models using only the green channel (GC Only).

\begin{table}[t]
    \caption{Performances  of base models with ablation.}\label{table_performance_base_models}
    \begin{tabularx}{\textwidth}{L{1.6}C{0.925}C{0.925}C{0.85}C{0.1}C{0.925}C{0.925}C{0.85}}
        \hline
         \multirow{2}{*}{Models} & \multicolumn{3}{c}{Cross. Point Val.} && \multicolumn{3}{c}{Severity Grade Pred.}\\
         \cline{2-4}\cline{6-8}
          &  Pre. & Rec. & $t$ (ms) && Acc. & Kappa & $t$ (ms)\\
        \hline
        ResNet-50 & 0.9427 & 0.9526 & 0.274 && 0.8063 & 0.6629 & 0.278\\
        \multicolumn{1}{l}{ \qquad---w/o PT} & 0.8646 & 0.6975 & 0.274 && 0.5445 & 0.0177 & 0.278\\
        \multicolumn{1}{l}{ \qquad---w/o DA} & 0.9531 & 0.8551 & 0.274 && 0.5340 & 0.0036 & 0.278\\
        \multicolumn{1}{l}{ \qquad---GC Only} & 0.9583 & 0.9154 & 0.273 && 0.7277 & 0.5288 & 0.273\\
        \hline
        Inception v3 & 0.9635 & \textbf{0.9635} & 0.218 && 0.8534 & 0.7432 & 0.222\\
        \multicolumn{1}{l}{ \qquad---w/o PT} & 0.9010 & 0.6865 & 0.218 && 0.5183 & 0.0313 & 0.222\\
        \multicolumn{1}{l}{ \qquad---w/o DA} & 0.9323 & 0.9179 & 0.218 && 0.5393 &  0.0000 & 0.222\\
        \multicolumn{1}{l}{ \qquad---GC Only} & 0.9167 & 0.9119 & \textbf{0.216} && 0.8115 & 0.6771 & \textbf{0.216}\\
        \hline
        DenseNet-121 & 0.9479 & 0.9630 & 0.266 && \textbf{0.8795} & \textbf{0.7892} & 0.269\\
        \multicolumn{1}{l}{ \qquad---w/o PT} & 0.9375 & 0.6742 & 0.266 && 0.5288 & 0.0050 & 0.269\\
        \multicolumn{1}{l}{ \qquad---w/o DA} & \textbf{0.9740} & 0.8274 & 0.266 && 0.7225 & 0.4865 & 0.269\\
        \multicolumn{1}{l}{ \qquad---GC Only} & \textbf{0.9740} & 0.9212 & 0.266 && 0.6702 & 0.4406 & 0.267\\
        \hline
    \end{tabularx}
\end{table}

\begin{table}[t]
    \caption{Performance of MDTNet models for severity grade prediction.}\label{table_performance_MDTNet}
    \begin{tabularx}{\textwidth}{L{1.1}C{1.1}C{1.1}C{1.1}C{1.1}C{0.2}C{1.1}C{1.1}C{1.1}}
        \hline
        \multirow{2}{*}{Metrics} & \multicolumn{4}{c}{DenseNet-121 (Focal Loss)} && \multicolumn{3}{c}{MDTNet}\\
         \cline{2-5}\cline{7-9} & $\gamma=1$ & $\gamma=2$ & $\gamma=5$ & $\gamma=10$ && $n=0$ & $n=1$ & $n=3$ \\
        \hline
        
        \hline
        Acc. & 0.8639 & 0.7434 & 0.8639 & 0.7958 && 0.8953 & 0.9110 & \textbf{0.9162}\\
        Kappa. & 0.7642 & 0.5685 & 0.7641 & 0.6508 && 0.8183 & 0.8453 & \textbf{0.8542}\\
        $t$ (ms) & \textbf{0.268} & \textbf{0.268} & \textbf{0.268} & \textbf{0.268} && 0.767 & 1.047 & 1.571\\
        \hline
    \end{tabularx}
\end{table}

\begin{figure}[t]
  \centering
  \subfloat[]{\includegraphics[width=0.335\textwidth]{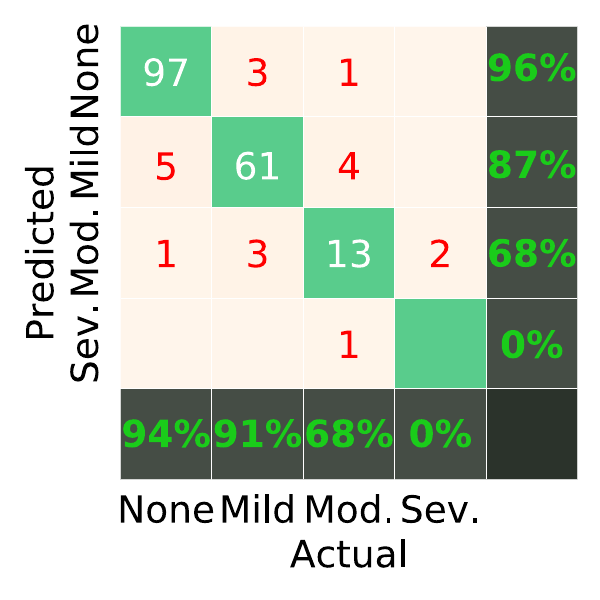}}
  \subfloat[]{\includegraphics[width=0.335\textwidth]{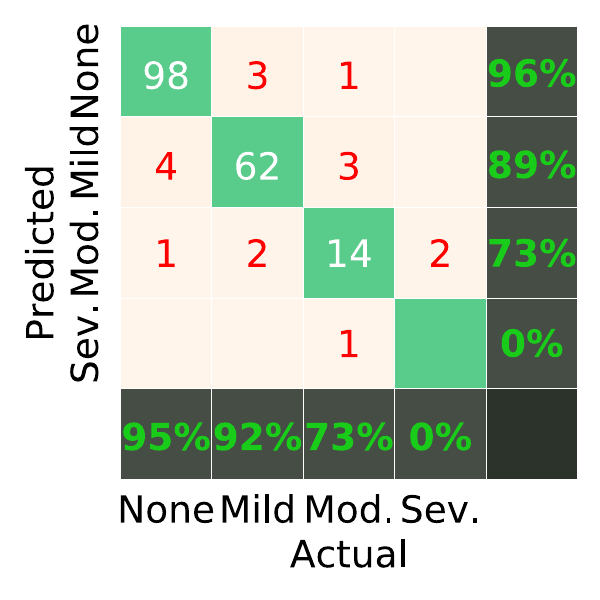}}
  \subfloat[]{\includegraphics[width=0.335\textwidth]{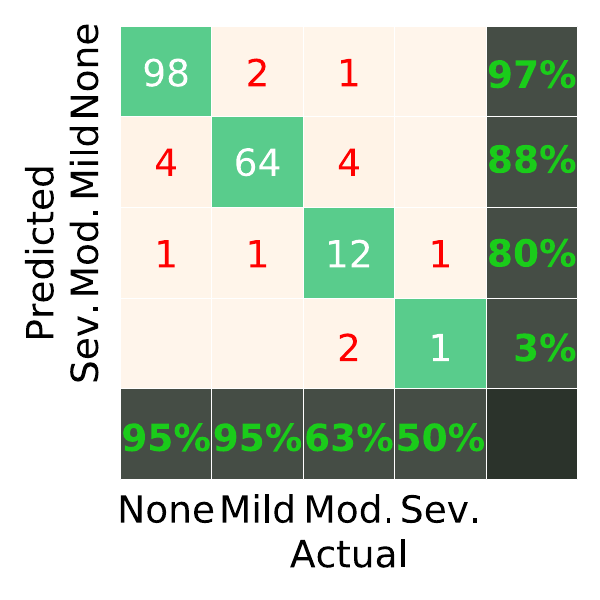}}
  \caption{Confusion matrices for three different severity grade prediction models. The recall is shown in the last row and the precision is shown in the last column. (a) MDTNet without the focal module, (b) MDTNet for $n=1$, and (c) MDTNet for $n=3$.}
  \label{fig_confusion_matrix}
\end{figure}

\begin{figure}[!t]
  \centering
  \includegraphics[width=1\textwidth]{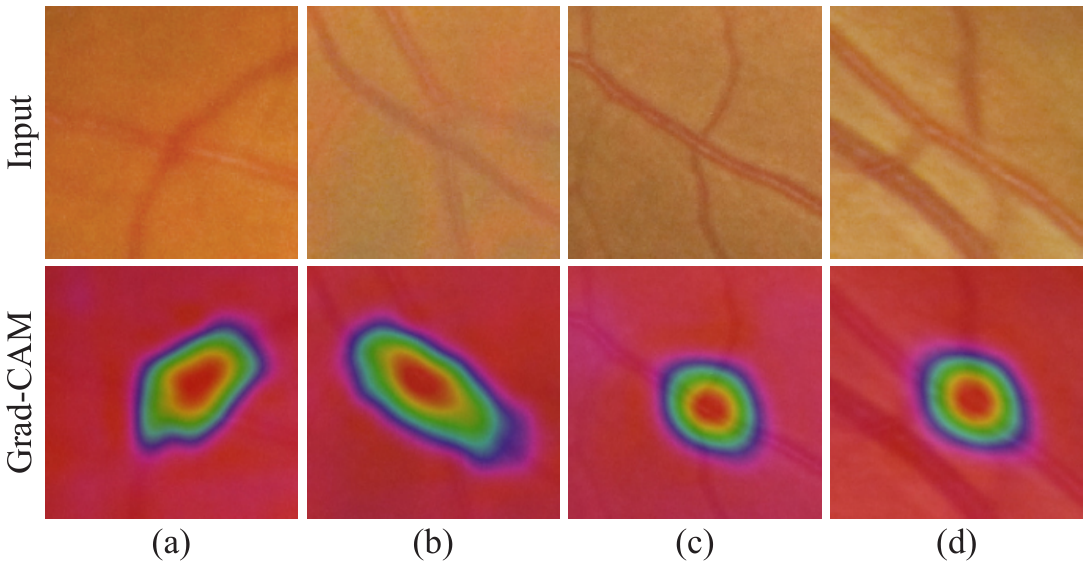}
  \caption{Visual explanation of prediction results. (a,b) are for the crossing point validation model and (c,d) are from the severity grade prediction model. The first row is the raw input images and the second row is the class-discriminative regions.}
  \label{fig_visual_explanation}
\end{figure}

The crossing point validation performances are shown in the left part of Table \ref{table_performance_base_models}. We use two metrics, precision and recall, and the time measurement to show the timing performance. We can see that pre-training and data augmentation can improve the overall performance of the crossing point validation. The Inception model with PT and DA achieved the best recall and the second-best precision.
Note that PT and DA will not change the running time of the model because they do not modify the network structure. 

The right part of Table \ref{table_performance_base_models} gives the results of the base models on the severity grade prediction task, and Table~\ref{table_performance_MDTNet} presents the performance of MDTNet and models using the focal loss. In addition to the classification accuracy, we also adopt \textcolor{black}{Cohen's kappa}, which can measure the agreement between the ground-truth labels and predictions. We can see that, compared with the focal loss models, the DenseNet can achieve higher overall accuracy with the CE loss. However, the combination \textcolor{black}{of} different models, different losses, as well as different $\gamma$ values can boost the performance. MDTNet achieved the highest performance in this experiment when $n=3$.


To better analyze the severity grade prediction performance, we present the confusion matrices in Fig.~\ref{fig_confusion_matrix}. 
It can be seen that, with the increment of the underlying sub-models, MDTNet gains the classification ability.
Fig.~\ref{fig_visual_explanation} shows visual explanation of MDTNet by Grad-CAM \cite{Grad-CAM}. Figs.~\ref{fig_visual_explanation} (a) and (b) show two examples for the crossing point validation. The ground-truth labels are \textit{false} and the predictions were also \textit{false}, \ie, these are not effective crossing points as the arteries are under the veins. The model mainly counted the red area in the second row along the vein. The model might find the vein, track it down, and \textcolor{black}{reach} the conclusion that it lies above the artery. Figs. \ref{fig_visual_explanation} (c) and (d) are for the severity grade prediction. The ground-truth labels are respectively \textit{mild} and \textit{moderate} and were both correctly predicted. We can see the artery runs over the vein deforming the vein. Being different from the example in (a) and (b), the model looks at the crossing points and looks for possible shape deformations and their extent.

\section*{Conclusion}
The paper presents a method to automatically classify the arteriolosclerosis severity from retinal images following ophthalmologists' diagnostic process. 
To improve the accuracy for ambiguous and unbalanced samples, we design the multi-diagnosis team network (MDTNet), which can jointly consider diagnostic cues from multiple sub-models, without tuning the hyperparameter for the focal loss. Experimental results show the superiority of our method, achieving over 91\% accuracy. Most importantly, the whole process can be checked to see how the grading was determined as \textcolor{black}{it is designed to be a step-by-step approach replicating ophthalmologists' diagnostic process}. Therefore, \textcolor{black}{the proposed method can serve as a supporting tool for experienced ophthalmologists to efficiently grade the images in a consistently reproducible manner.}
\textcolor{black}{A quality checklist \cite{norgeot2020minimum} for the proposed deep learning method is shown in Table.~\ref{table_ai_checklist}.}

\section*{Author Contributions}
\paragraph*{Conceptualization}: Liangzhi Li, Yuta Nakashima, Hajime Nagahara, Ryo Kawasaki
\paragraph*{Data curation}: Liangzhi Li, Bowen Wang
\paragraph*{Formal analysis}: Liangzhi Li, Manisha Verma
\paragraph*{Funding acquisition}: Liangzhi Li, Ryo Kawasaki
\paragraph*{Methodology}: Liangzhi Li, Manisha Verma
\paragraph*{Project administration}: Yuta Nakashima, Hajime Nagahara, Ryo Kawasaki
\paragraph*{Supervision}: Yuta Nakashima, Hajime Nagahara, Ryo Kawasaki
\paragraph*{Validation}: Liangzhi Li, Bowen Wang
\paragraph*{Writing-original draft}: Liangzhi Li, Manisha Verma
\paragraph*{Writing-review \& editing}: Liangzhi Li, Manisha Verma, Yuta Nakashima, Ryo Kawasaki

\begin{table}
    \caption{\textcolor{black}{The MI-CLAIM checklist.}}
    \label{table_ai_checklist}
    \begin{tabular}{|p{0.6\textwidth}|>{\centering\arraybackslash}p{0.1\textwidth}|>{\centering\arraybackslash}p{0.1\textwidth}|}
    \hline
        
\multicolumn{3}{|l|}{Before paper submission}
 \\ \hline
Study design (Part 1)
 & 
Page
 & 
Notes
 \\ \hline
The clinical problem in which the model will be employed is clearly detailed in the paper.
 & 2-3 &  \\ \hline
The research question is clearly stated.
 & 3 &  \\ \hline
The characteristics of the cohorts (training and test sets) are detailed in the text.
 & 3-4 &  \\ \hline
The cohorts (training and test sets) are shown to be representative of real-world clinical settings.
 & 3-4 &  \\ \hline
The state-of-the-art solution used as a baseline for comparison has been identified and detailed.
 & 7-8 &  \\ \hline
Data and optimization
 & 
Page
 & 
Notes
 \\ \hline
The origin of the data is described and the original format is detailed in the paper.
 & 3-4 &  \\ \hline
Transformations of the data before it is applied to the proposed model are described.
 & 6-7 &  \\ \hline
The independence between training and test sets has been proven in the paper.
 & 4 &  \\ \hline
Details on the models that were evaluated and the code developed to select the best model are provided.
 & 7 &  \\ \hline
Is the input data type structured or unstructured?
 & 
 \multicolumn{2}{|c|}{
 \makebox[0pt][l]{$\square$}\raisebox{.15ex}{\hspace{0.1em}$\checkmark$} Structured \quad 
 \makebox[0pt][l]{$\square$}\quad 
Unstructured}
 \\ \hline
Model performance (Part 4)
 & 
Page
 & 
Notes
 \\ \hline
The primary metric selected to evaluate algorithm performance, including the justification for selection, has been clearly stated.
 & 7-8 &  \\ \hline
The primary metric selected to evaluate the clinical utility of the model, including the justification for selection, has been clearly stated.
 & 7-8 &  \\ \hline
The performance comparison between baseline and proposed model is presented with the appropriate statistical significance.
 & 7-9 &  \\ \hline
Model examination (Part 5)
 & 
Page
 & 
Notes
 \\ \hline
Examination technique
 & 7-9
 &  \\ \hline
A discussion of the relevance of the examination results with respect to model/algorithm performance is presented.
 & 7-9 &  \\ \hline
A discussion of the feasibility and significance of model interpretability at the case level if examination methods are uninterpretable is presented.
 & 8 &  \\ \hline
A discussion of the reliability and robustness of the model as the underlying data distribution shifts is included.
 & 4,7-9 &  \\ \hline
\multicolumn{2}{|l|}{Reproducibility (Part 6): choose appropriate tier of transparency}
 & 
Notes
 \\ \hline
 Tier 1: complete sharing of the code
 & 
\quad\makebox[0pt][l]{$\square$}\raisebox{.15ex}{\hspace{0.1em}$\checkmark$}
 &  \\ \hline
Tier 2: allow a third party to evaluate the code for accuracy/fairness; share the results of this evaluation
 & 
\makebox[0pt][l]{$\square$}
 &  \\ \hline
Tier 3: release of a virtual machine (binary) for running the code on new data without sharing its details
 & 
\makebox[0pt][l]{$\square$}
 &  \\ \hline
Tier 4: no sharing
 & 
\makebox[0pt][l]{$\square$}
 &  \\ \hline
    \end{tabular}
\end{table}


%
%
%

\newpage

\begin{thebibliography}{10}

\bibitem{chatziralli2012value}
Chatziralli IP, Kanonidou ED, Keryttopoulos P, Dimitriadis P, Papazisis LE.
\newblock The value of fundoscopy in general practice.
\newblock The open ophthalmology journal. 2012;6:4.

\bibitem{hubbard1999methods}
Hubbard LD, Brothers RJ, King WN, Clegg LX, Klein R, Cooper LS, et~al.
\newblock Methods for evaluation of retinal microvascular abnormalities
  associated with hypertension/sclerosis in the Atherosclerosis Risk in
  Communities Study.
\newblock Ophthalmology. 1999;106(12):2269--2280.

\bibitem{WALSH19821127}
Walsh JB.
\newblock Hypertensive Retinopathy: Description, Classification, and Prognosis.
\newblock Ophthalmology. 1982;89(10):1127 -- 1131.

\bibitem{6547196}
{Nguyen} UTV, {Bhuiyan} A, {Park} LAF, {Kawasaki} R, {Wong} TY, {Wang} JJ,
  et~al.
\newblock An Automated Method for Retinal Arteriovenous Nicking Quantification
  From Color Fundus Images.
\newblock IEEE Transactions on Biomedical Engineering. 2013;60(11):3194--3203.

\bibitem{7042289}
{Roychowdhury} S, {Koozekanani} DD, {Parhi} KK.
\newblock Iterative Vessel Segmentation of Fundus Images.
\newblock IEEE Transactions on Biomedical Engineering. 2015;62(7):1738--1749.

\bibitem{8036917}
{Kim} JU, {Kim} HG, {Ro} YM.
\newblock Iterative deep convolutional encoder-decoder network for medical
  image segmentation.
\newblock In: IEEE Engineering in Medicine and Biology Society (EMBC); 2017. p.
  685--688.

\bibitem{8341481}
{Yan} Z, {Yang} X, {Cheng} K.
\newblock Joint Segment-Level and Pixel-Wise Losses for Deep Learning Based
  Retinal Vessel Segmentation.
\newblock IEEE Transactions on Biomedical Engineering. 2018;65(9):1912--1923.

\bibitem{HUANG2018197}
Huang F, Dashtbozorg B, Tan T, ter Haar~Romeny BM.
\newblock Retinal artery/vein classification using genetic-search feature
  selection.
\newblock Computer Methods and Programs in Biomedicine. 2018;161:197 -- 207.

\bibitem{10.1007/978-3-319-93000-8_71}
Meyer MI, Galdran A, Costa P, Mendon{\c{c}}a AM, Campilho A.
\newblock Deep Convolutional Artery/Vein Classification of Retinal Vessels.
\newblock In: Image Analysis and Recognition; 2018. p. 622--630.

\bibitem{8055572}
{Costa} P, {Galdran} A, {Meyer} MI, {Niemeijer} M, {Abràmoff} M, {Mendonça}
  AM, et~al.
\newblock End-to-End Adversarial Retinal Image Synthesis.
\newblock IEEE Transactions on Medical Imaging. 2018;37(3):781--791.

\bibitem{hatanaka2011automatic}
Hatanaka Y, Muramatsu C, Hara T, Fujita H.
\newblock Automatic arteriovenous crossing phenomenon detection on retinal
  fundus images.
\newblock In: Medical Imaging 2011: Computer-Aided Diagnosis. vol. 7963; 2011.
  p. 79633V.

\bibitem{li2019iternet}
Li L, Verma M, Nakashima Y, Nagahara H, Kawasaki R.
\newblock {IterNet}: Retinal Image Segmentation Utilizing Structural Redundancy
  in Vessel Networks.
\newblock In: The IEEE Winter Conference on Applications of Computer Vision;
  2020. p. 3656--3665.

\bibitem{cambocombined}
Camb{\`o} VBS, Cariello L, Mastronardi G.
\newblock A COMBINED METHOD TO DETECT RETINAL FUNDUS FEATURES.
\newblock In: IEEE European Conference on Emergent Aspects in Clinical Data
  Analysis; 2005.

\bibitem{taylor2010multidisciplinary}
Taylor C, Munro AJ, Glynne-Jones R, Griffith C, Trevatt P, Richards M, et~al.
\newblock Multidisciplinary team working in cancer: what is the evidence?
\newblock {The BMJ}. 2010;340:c951.

\bibitem{FocalLoss}
{Lin} T, {Goyal} P, {Girshick} R, {He} K, {Dollár} P.
\newblock Focal Loss for Dense Object Detection.
\newblock IEEE Transactions on Pattern Analysis and Machine Intelligence.
  2020;42(2):318--327.

\bibitem{inoue2009stroke}
Inoue R, Ohkubo T, Kikuya M, Metoki H, Asayama K, Kanno A, et~al.
\newblock Stroke risk of blood pressure indices determined by home blood
  pressure measurement: the Ohasama study.
\newblock Stroke. 2009;40(8):2859--2861.

\bibitem{li2020joint}
Li L, Verma M, Nakashima Y, Kawasaki R, Nagahara H.
\newblock Joint Learning of Vessel Segmentation and Artery/Vein Classification
  with Post-processing.
\newblock In: Medical Imaging with Deep Learning; 2020.

\bibitem{resnet}
{He} K, {Zhang} X, {Ren} S, {Sun} J.
\newblock Deep Residual Learning for Image Recognition.
\newblock In: The IEEE Conference on Computer Vision and Pattern Recognition
  (CVPR); 2016. p. 770--778.

\bibitem{densenet}
Huang G, Liu Z, Pleiss G, Van Der~Maaten L, Weinberger K.
\newblock Convolutional Networks with Dense Connectivity.
\newblock IEEE Transactions on Pattern Analysis and Machine Intelligence.
  2019;.

\bibitem{inception}
{Szegedy} C, {Vanhoucke} V, {Ioffe} S, {Shlens} J, {Wojna} Z.
\newblock Rethinking the Inception Architecture for Computer Vision.
\newblock In: The IEEE Conference on Computer Vision and Pattern Recognition
  (CVPR); 2016. p. 2818--2826.

\bibitem{AutomatedFocalLoss}
Weber M, F{\"u}rst M, Z{\"o}llner JM.
\newblock Automated Focal Loss for Image based Object Detection.
\newblock arXiv preprint arXiv:190409048. 2019;.

\bibitem{7780949}
{Huang} C, {Li} Y, {Loy} CC, {Tang} X.
\newblock Learning Deep Representation for Imbalanced Classification.
\newblock In: 2016 IEEE Conference on Computer Vision and Pattern Recognition
  (CVPR); 2016. p. 5375--5384.

\bibitem{Cui_2019_CVPR}
Cui Y, Jia M, Lin TY, Song Y, Belongie S.
\newblock Class-Balanced Loss Based on Effective Number of Samples.
\newblock In: The IEEE Conference on Computer Vision and Pattern Recognition
  (CVPR); 2019.

\bibitem{ILSVRC15}
Russakovsky O, Deng J, Su H, Krause J, Satheesh S, Ma S, et~al.
\newblock {ImageNet Large Scale Visual Recognition Challenge}.
\newblock International Journal of Computer Vision (IJCV).
  2015;115(3):211--252.

\bibitem{adam}
Kingma DP, Ba J.
\newblock Adam: A method for stochastic optimization.
\newblock arXiv preprint arXiv:14126980. 2014;.

\bibitem{Grad-CAM}
{Selvaraju} RR, {Cogswell} M, {Das} A, {Vedantam} R, {Parikh} D, {Batra} D.
\newblock Grad-CAM: Visual Explanations from Deep Networks via Gradient-Based
  Localization.
\newblock In: International Conference on Computer Vision (ICCV); 2017. p.
  618--626.

\bibitem{norgeot2020minimum}
Norgeot B, Quer G, Beaulieu-Jones BK, Torkamani A, Dias R, Gianfrancesco M,
  et~al.
\newblock Minimum information about clinical artificial intelligence modeling:
  the MI-CLAIM checklist.
\newblock Nature medicine. 2020;26(9):1320--1324.

\end{thebibliography}

\end{document}